\pdfoutput=1

\documentclass[conference]{IEEEtran}
\ifCLASSINFOpdf
\else
\fi
\usepackage{etex}

\usepackage[]{graphicx}
\usepackage[cmex10]{amsmath}
\usepackage{amssymb}
\usepackage{enumerate}
\usepackage{booktabs}
\usepackage{tabulary}
\usepackage{colortbl}
\usepackage{color}
\usepackage{eqnarray}
\usepackage{amsmath}  
\usepackage[usenames,dvipsnames,svgnames,table]{xcolor}
\usepackage{mathtools}
\usepackage{adjustbox}
\usepackage{tabularx}
\usepackage{colortbl}

\usepackage{amsthm}

\theoremstyle{definition}

\usepackage{multirow}

\usepackage[disable,colorinlistoftodos,prependcaption,textsize=tiny]{todonotes}
\usepackage{lipsum}                     
\usepackage{xargs}                      
\usepackage{subfig} 
\usepackage{epstopdf}

\usepackage[colorinlistoftodos,prependcaption,textsize=tiny]{todonotes}
\newcommandx{\unsure}[2][1=]{\todo[linecolor=red,backgroundcolor=red!25,bordercolor=red,#1]{#2}}
\newcommandx{\change}[2][1=]{\todo[linecolor=blue,backgroundcolor=blue!25,bordercolor=blue,#1]{#2}}
\newcommandx{\info}[2][1=]{\todo[linecolor=OliveGreen,backgroundcolor=OliveGreen!25,bordercolor=OliveGreen,#1]{#2}}
\newcommandx{\improvement}[2][1=]{\todo[linecolor=Plum,backgroundcolor=Plum!25,bordercolor=Plum,#1]{#2}}
\newcommandx{\cmnt}[2][1=]{\todo[linecolor=Blue,backgroundcolor=Blue!25,bordercolor=Blue,#1]{#2}}
\newcommandx{\thiswillnotshow}[2][1=]{\todo[disable,#1]{#2}}
\begin{document}

%

\title{\vspace{-10pt}
	Bandwidth management VMs live migration \\ in wireless fog computing for 5G networks\vspace{-10pt}}

\author{\IEEEauthorblockN{Danilo Amendola}
\IEEEauthorblockA{DIET, Sapienza University of Rome\\
danilo.amendola@uniroma1.it}
\and
\IEEEauthorblockN{Nicola Cordeschi}
\IEEEauthorblockA{DIET, Sapienza University of Rome\\
nicola.cordeschi@uniroma1.it}
\and
\IEEEauthorblockN{Enzo Baccarelli}
\IEEEauthorblockA{DIET, Sapienza University of Rome\\
enzo.baccarelli@uniroma1.it}
}



\maketitle


\begin{abstract}
Live virtual machine migration aims at enabling the dynamic balanced use of the networking/computing physical resources of virtualized data-centers, so to lead to reduced energy consumption. Here, we analytically characterize, prototype in software and test an optimal bandwidth manager for live migration of VMs in wireless channel.
In this paper we present the optimal tunable-complexity bandwidth manager (TCBM) for the QoS live migration of VMs under a wireless channel from smartphone to access point. The goal is the minimization of the migration-induced communication energy under service level agreement (SLA)-induced \emph{hard} constrains on the total migration time, downtime  and overall available bandwidth.
\improvement[inline]{Abstract da completare.}
\end{abstract}


%
\IEEEpeerreviewmaketitle

\section{Introduction and related work}

Mobile cloud computing (MCC) emerging in the context of 5G has the potential to overcome resource limitation in the mobile devices (appear as a bottleneck in 5G applications), which enables many resource-intensive services for mobile users with the support of mobile big data delivery and cloud-assisted computing \cite{Chen2015}.

In 5G a fundamental issue is to provide services with low latency Fog computing (FC), also know as edge computing, can address those problems by providing elastic resources and services to end users at the edge of the network.
The difference between fog computing and cloud computing (CC) is that CC focuses on providing resources located in the core network, while FC focuses on resources distributed in the edge network.

In this context a plethora of frameworks and models (oriented to the MCC) are proposed, to provide high performance computation system on mobile devices. 
We briefly present in the follows some of these solutions:

\begin{itemize}
\item CloneCloud \cite{clonecloudchun2011,schuring2011mobile}: is a system that has the ability to automatically transform mobile device application in such a way that they can run into the cloud;
\item VOLARE \cite{volarepapakos2010}: is a middelware-based solution which allows context-aware adaptive cloud service discovery for the mobile devices.
\item Cuckoo \cite{cuckookemp2010}: is a computational offloading framework for mobile devices;
\item Cloudlet \cite{cloudlet2009case}: is a set of widely dispersed and decentralized Internet infrastructure components, with non-trivial characteristic to make available for the nearby mobile devices computing resource and storage resources;
\item MAUI \cite{mauicuervo2010}: is a system that is able to minimize the energy due to the VM migration by means of fine-grained offloading.
\end{itemize} 

This paper is organized as follows. Section \ref{sec:livemig} gives a brief description of the live migration problem. Section \ref{sec:tcbm} introduces our bandwidth manager, formulation and solution of the non-convex optimization problem.  Section \ref{sec:experiment} shows experimental work and tests. Finally, we conclude our work in Section \ref{sec:conclusion}.

\section{The Tackled problem: Live Migration}
\label{sec:livemig}
Virtualization is a viral technology in the data center and hardware efficient utilization, its benefit is well recognized in a large number of applications. Virtualization \cite{barham2003xen} is rapidly evolving and live migration is a core function to replace running VMs seamlessly across distinct physical devices \cite{7-Clark}.

In recent years considerable interest has been pointed out on VM live migration for data center migration \cite{7-Clark}  and cluster computing\improvement{[cit altri?]}.

Efficient VM live migration is an attractive function in virtualized systems cause this is essential to enable consolidation techniques oriented to save energy consumption. Representative technologies for VM live migration are XenMotion \cite{7-Clark} and VMware VMotion, both of them implemented as a built-in tool in their virtualized platforms. There are also other studies about VM migration in which the problem of where and when a VM should be migrated \improvement[]{[cit circa consolidation e migr dove e quando???]} to improve the system performances is considered. But none of them are considering the issue of how to improve the communication performance with bandwidth optimization for migration when time and place of migration are decided.

Then VM live migration technologies are very effective tool to enable data-center management and save energy consumption. 
During the live migration, physical memory image is transferred across the network to the new destination, while the source VM continue to run until the last bit will be received with success.
We treated this issue in our previous work \cite{Baccarelli20151}, we considered the intra data-center channel optimization bandwidth problem. Hence, here we investigate live virtual machine migration bandwidth optimization on wireless channel. Besides, these works \cite{baccarelli2005broadband, cordeschi2012stochastic, cordeschiCCNC15, Cordeschi-DSRT} are useful to understand our approach.

In literature there are four main techniques for VM migration, namely, stop-and-copy migration (SaCM), pre-copy migration (PeCM), post-copy migration (PoCM) and hybrid migration (HyBM). They trade-off the total migration time and downtime. 
Here, to be concise, we omitted a complete overview of main techniques for VM live migration. To understand how it works you can refer to our work \cite{Baccarelli20151}. In the following we use the pre-copy live migration technique, as in \cite{Baccarelli20151}. Our approach may be applied to all the mentioned techniques.

Considering the related work, at this time there are not works considering the bandwidth management during the VMs live migration for wireless channel. Our previous work \cite{Baccarelli20151} is the first which considers the bandwidth management in wired network environment. In that work we presented a bandwidth manager atop an intra-data-center wired test-bed comparing performances with most relevant VMs live technologies.

As we described in \cite{Baccarelli20151}, this approach is capable to effectively filter out transient fluctuations of the average resource utilization and avoid needless migrations \cite{14-Wood}.

\section{Tunable Complexity Bandwidth Management definition and basic properties}
\label{sec:tcbm}
In this section we introduce the tunable complexity bandwidth management (TCBM).
Let $I_{max}$ be the number of performed pre-copy rounds. 

\begin{figure}[!h]
	\centering
	\includegraphics[width=1.0\columnwidth,natwidth=5896,natheight=4026]{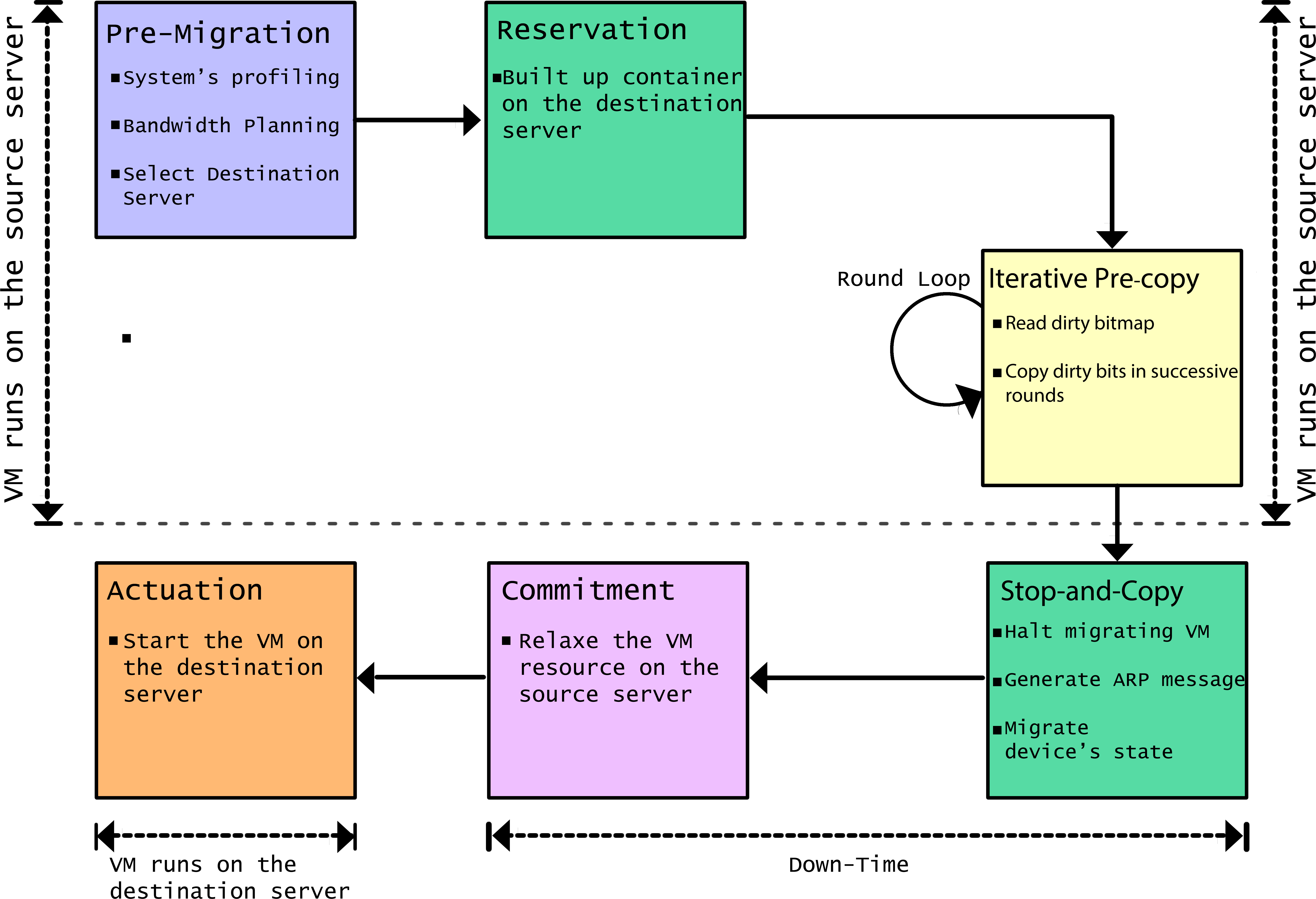}
	\caption{Pre-copy live migration stages (six stages).}
	\label{fig:stage_techniques}
\end{figure}

A primary goal of our work is to formal define a model overview of how live migration works. Most important variables are total migration time and downtime.
From a formal point of view, the total migration time $T_{TOT}$ $(s)$ is the overall duration:
	$T_{TOT} \triangleq  T_{PM}+T_{RE}+T_{IP}+T_{SC}+T_{CM}+T_{AT}$,
of the six stages (as we can see in Fig. \ref{fig:stage_techniques}), while the downtime: 
	$T_{DT} \triangleq T_{SC}+T_{CM}+T_{AT}$, 
is the time required for the execution of the last three stages. From a practical point of view, $T_{TOT}$ is the period when the states of the source and destination servers must be synchronized, which may also affect the reliability of the migration process, 
while ${T_{DT}}$ 
is the period in which the migrating VM is halted and the clients experience a service outage \cite{2-Xu}.

Let $R_i$ $(Mb/s)$ be the transmission rate used during the third and fourth stages at the $ith$ round for migrating the VM, that is, the migration bandwidth. However, we present a first formulation of the problem considering a constant rate at each round, $R_i = R \: \forall i$. Since, by definition, only ${T_{IP}}$ and ${T_{SC}}$ depend 
on $R$, while all the remaining migration times in $T_{TOT}$ and $T_{DT}$ 
play the role of constant parameters, in the sequel, we focus on the evaluation of the (already defined) stop-and-copy time ${T_{SC}}$ and the resulting memory migration time $T_{MMT}$, which is defined as in:
\begin{equation}
	\label{eqn_3}
	T_{MMT}~ \equiv~ T_{MMT}(R) ~\triangleq~ T_{IP}(R)+T_{SC}(R).
\end{equation}
Table \ref{table:1} reports the definitions of the key parameters used in the paper. Since the PeCM technique performs the iterative pre-copy of dirtied memory bits over consecutive rounds, let $V_{i}$ $(Mb)$ and $T_{i}$ $(s)$, $i~=~0, \ldots, ({I_{MAX}+1})$, be the volume of the migrated data and the time duration of the $ith$ round, respectively. By definition, $V_{0}$ and $T_{0}$ are the memory size $M_{0}$  $(Mb)$ of the migrating VM and the time needed for migrating it during the $0th$ round, respectively. 

\begin{table}[!h]
	\centering
	\caption{Main taxonomy of the paper.}
	\begin{tabular}{l l r}
	\cmidrule{0-1}
	\textbf{Symbol} & \textbf{Meaning/Role} \\
	\hline \hline
	\rowcolor[gray]{0.9}$I_{MAX}$ & Number of migration pre${-}$copy rounds \\
	\hline
	$i$ & Round index, $i {=}0 $,$\ldots$,(${I_{MAX}+1}$) \\
	\hline
	\rowcolor[gray]{0.9} $\overline{\textit{w}} {(Mb/s)}$ & Memory dirty rate of the migrated VM\\
	\hline
	$\hat{R_i}$ ${(Mb/s)}$ & Migration bandwidth at $ith$ round\\
	\hline
	\rowcolor[gray]{0.9} $P(R_i)$ $(W)$ & Communication power at the migration bandwidth $R_i$\\
	\hline
	$\hat{R}$ ${(Mb/s)}$ & Maximum available migration bandwidth\\
	\hline
	\rowcolor[gray]{0.9} $M_{0}$ $(Mb)$ & Memory size of the migrated VM\\
	\hline
	$\mathcal{E}_{TOT}$ $(J)$ & Total consumed communication energy\\
	\hline
	\rowcolor[gray]{0.9}$\Delta_{MMT}$ $(s)$ & Maximum tolerated memory migration time\\
	\hline
	$\Delta_{SC}$ $(s)$ & Maximum tolerated		stop${-}$and${-}$copy time\\
	\hline
	\rowcolor[gray]{0.9}$\beta$ & Migration speed${-}$up factor\\
	\hline
	$n$ & Integer${-}$valued iteration index\\
	\hline
	\cmidrule{0-1}
	\end{tabular}
	\label{table:1}
\end{table}

Now we formalize the afforded tunable-complexity bandwidth manager. In addition to $R_{0}$ and $R_{I_{MAX}+1}$ we have $Q$, which is the number of updated rates. Then we updated $Q$ out of $I_{MAX}$ rates of the pre-copy rounds evenly spaced by $S \triangleq \dfrac{I_{MAX}}{Q}$ over the round-index set \{1,2,3,\dots,$I_{MAX}$\}.

For this purpose, we perform the partition of the round index set \{1,2,3,\dots,$I_{MAX}$\} into Q not overlapping contiguous subsets of size $S$. 

\begin{figure}[h]
	\begin{center}
		\includegraphics[width=1.0\columnwidth,natwidth=3000,natheight=1000]{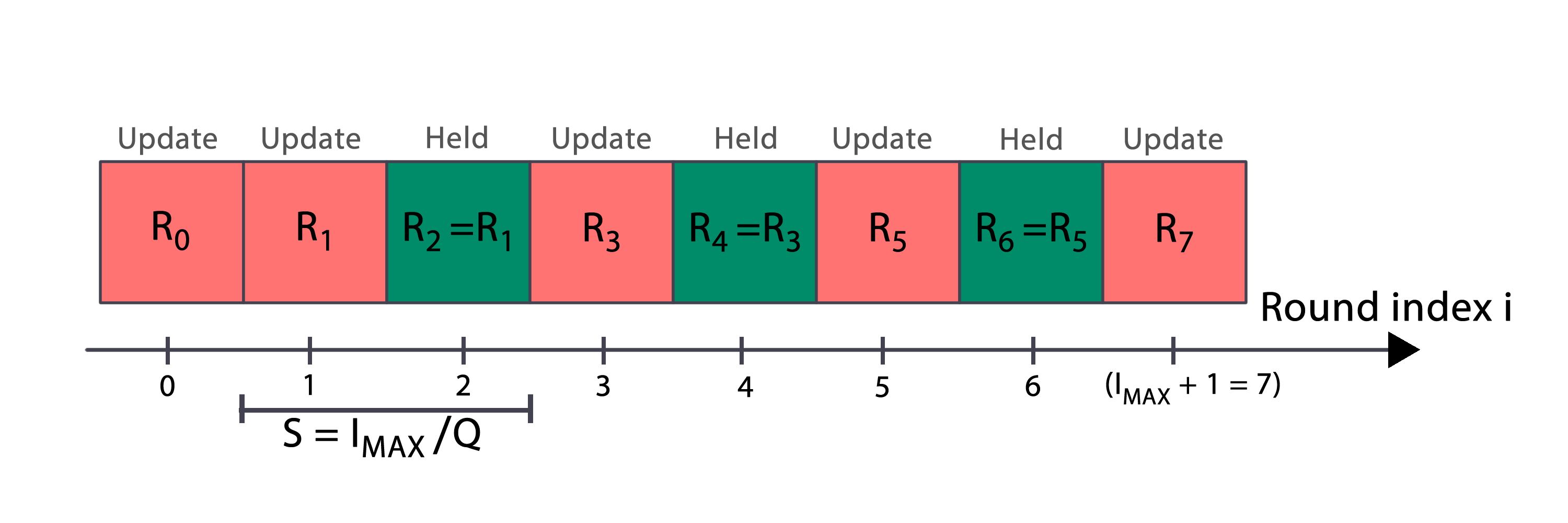}
		\caption{Reference framework for the tunable-complexity bandwidth manager. Case of $I_{MAX} = 6$ , $Q = 3$. The rates to be uploaded are: $R_{0},R_{1},R_{3},R_{5}$ and $R_{7}$. The rates to be held are: $R_{2} \equiv R_{1}$; $R_{4} \equiv R_{3}$; $R_{6} \equiv R_{5}$.}
		\label{fig:figTunable} 
	\end{center}
	\vspace{-10pt}
\end{figure}

The fist rate $R_{jS + 1}$ , $j = 0,\dots,(Q-1)$ of each subset is updated, while the remaining $(S - 1)$ rates are set to $R_{jS + 1}$, that is $R_{i}\equiv R_{jS + 1}$, for $i = jS+2$, $jS+3$, \dots, $(j + 1)S$.

Fig. \ref{fig:figTunable} illustrates the framework of the updated/held migration rates for the dummy case of $I_{MAX} = 6$ and $Q = 3$. In this case, $R_{0}, R_{1}, R_{3}, R_{5}$ and $R_{7}$ are the Q + 2 = 5 migration rates to be updated, while $R_{2}, R_{4}$ and $R_{6}$ are the $(I_{MAX} - Q) = 3$ migration rates which are not updated and, by definition, they equate: $R_{2} \equiv R_{1};\: R_{4}\equiv R_{3};\: R_{6}\equiv R_{5}$.

\subsection{Formulation of the non-convex optimization problem to be solved by  the TCBM}
\label{sez:3.4.6}
The TCBM is the solution of the following non-convex optimization problem, which could be solved as an instance of geometric problem (solution is omitted here for briefness, see \cite{Baccarelli20151} for details):

\begin{equation}\label{eq:n38}
\min_{	\{ R_{0},R_{jS+1}, \:j = 0, 1, \dots, (Q-1); \:R_{I_{MAX}+1}\}} \mathcal{E}_{TOT}
\end{equation}
s.t. 
\begin{equation}\label{eq:n39}
\small
\Psi_{1}  \triangleq\theta \biggl\{  \biggl( \dfrac{1}{\Delta}_{TM} T_{TM} \biggl) -1 \biggl\}\leq0;
\end{equation}
\begin{equation}\label{eq:n40}
\small
\Psi_{2}  \triangleq\biggl( \dfrac{1}{\Delta}_{DT}T_{DT} \biggl) -1 \leq 0; 
\end{equation}
\begin{equation}\label{eq:n41}
\small
\begin{split}
\Psi_{3}  \triangleq\theta \: \biggl\{ \beta\: \overline{w} \:R_{i}^{-1} - 1\biggl\} \leq 0, \\
for\: i=0;\: i=jS+1;\: j= 0, \dots, (Q-1);
\end{split}
\end{equation}
\begin{equation}\label{eq:n42}
\small
\centering
\begin{split}
R_{i} \leq \widehat{R},\\ 
for\: i=0;\: i=jS+1;\: j= 0,\: \dots, (Q-1); \: i= I_{MAX} + 1;
\end{split}
\end{equation}

Four constraints are considered in the formulation of the TCBM, which capture, in turn, the metrics currently adopted for measuring the performance of live migration techniques \cite{19a_xu2014managing,33a_wu2011performance}. The first two constraints upper limit the tolerated total migration time and downtime. Constrain \eqref{eq:n41} account the ratio of the volumes of data migrated over two consecutive rounds falls below a predefined speed-factor $\beta > 1$. Finally, constrain \eqref{eq:n42} upper limit the maximum available rate. Furthermore, the $\theta$ parameter in \eqref{eq:n39} accounts for the fact that, by definition, the total migration and stop-and-copy times coincide under the SaCM and PoCM techniques.

\section{Experimental work and tests}
\label{sec:experiment}

In order to actually test and compare the performance of the proposed bandwidth manager, we have implemented an experimental wireless test-bed.

Below we discuss some experiments that show the goodness of our TCBM, comparing with the results obtained from Xen and the method BMOP (Bandwidth Management Optimization Problem, see \cite{Baccarelli20151} for implementation) in which, unlike in our software, the initial rate, is held for the entire duration of the VM migration . 

Of practical interest, specifically, the reported data refer to the average parameters of typical wireless IEEE 802.11b, 3G-UTRAN and 4G-LTE connections. We anticipated that the reported data are in agreement with \cite{43a_perrucci2011survey} for 3G-UTRAN and \cite{44a_huang2012close} for 4G-LTE. 

After noting that $\tilde{I}_{MAX}$ refers to our optimized setting of the allowed pre-copy rounds, typically values for the tested VMs are: $1 \: \leq\: \tilde{I}_{MAX}\: \leq \: 29$, 
where $\tilde{I}_{MAX} = 29$ is the Xen's default setting;
$R_{MAX}$ is: $0.9 \times 2 (Mb/s)$ for 3G cellular; $0.9 \times 11 (Mb/s)$ for IEEE 802.11b; and  $0.9 \times 50 (Mb/s)$ for 4G-LTE, 
where $R_{MAX}$ (Mb/s) is the maximum throughput at the Transport Layer. 
${E}_{SETUP}$ is: $3.25 (J)$ for 3G cellular; $5.9 (J)$ for IEEE 802.11.b; $5.1 (J)$ for 4G-LTE.
where $\mathcal{E}_{SETUP}$ is the static (e.g., rate independent) part of the overall energy consumption of the considered connection.

All tests have been carried out in three different application scenarios, i.e., the scenario in which the smartphone migrates to the access point by 3g; the scenario in which the smartphone migrates with the use of the 4G; and finally the scenario where migration is performed by making use of WiFi.
	
\subsection{The benchmark Xen bandwidth management}
\label{sez:4.2} 
The currently implemented Xen hypervisor adopts a pre-copy heuristic bandwidth management policy, which operates on a best effort basis, while attempting to shorten the final stop-and-copy time \cite{46a_hwang2013distributed, 47a_clark2005live}. The rationale behind this Xen policy is that, in principle, the stop-and-copy time may be reduced by monotonically increasing the migration bandwidth over consecutive rounds \cite{47a_clark2005live}.  For this purpose, the Xen hypervisor uses pre-assigned minimum: $R_{MIN}^{XEN}$ (Mb/s), and maximum:
$R_{MAX}^{XEN}$ $(Mb/s)$ bandwidth thresholds, in order to bound the migration bandwidth during the pre-copy stage (see Section 5.3 of \cite{47a_clark2005live}). Specifically, the Xen migration bandwidth $R^{XEN}$ equates: $R_{MIN}^{XEN}$ (Mb/s) at round\#0, and, then, it increases in each subsequent round by a constant term:  $\Delta R^{XEN}$ $(Mb/s)$, so to reach the maximum value: $R^{XEN} = R_{MAX}^{XEN}$ at the last round: round\#($I_{MAX}$ + 1) (see Section 5.3 of \cite{47a_clark2005live}). In the carried out field trials, we have implemented this benchmark policy by setting:

\begin{equation}\label{eq:n4.10}
\Delta R^{XEN} = (R_{MAX}^{XEN} \:-\: \overline{w}) \: / \: (I_{MAX}^{XEN} + 1),
\end{equation}
and
\begin{equation}\label{eq:n4.11}
R_{i}^{XEN} = \overline{w} \: + \: i \Delta R^{XEN}, \:\:\:\: i = 0, \dots, (I_{MAX}^{XEN} + 1).
\end{equation}
We point out that, on the basis of the (recent) surveys in \cite{19a_xu2014managing}, Chapter 3 of \cite{46a_hwang2013distributed} and Chapter 17 of \cite{48a_commInfra}, this is the only bandwidth management policy currently considered by both academy and industry for VM migration. This is also the bandwidth policy currently implemented by Xen, KVM and VMware commercial hypervisors \cite{46a_hwang2013distributed}.

\subsection{Tracking capabilities under contention phenomena}
\label{sez:4.4}
Real-world applications may vary the produced traffics over the time \cite{cordeschi2015energy} and, then, it may be of interest to test how the proposed bandwidth manager reacts when the workload offered by the migrating VM changes unexpectedly. 

\begin{figure*}[]
\subfloat[{$\overline{w} = [0.8, 1.5, 0.8]$} ]{\includegraphics[width=.6\columnwidth]{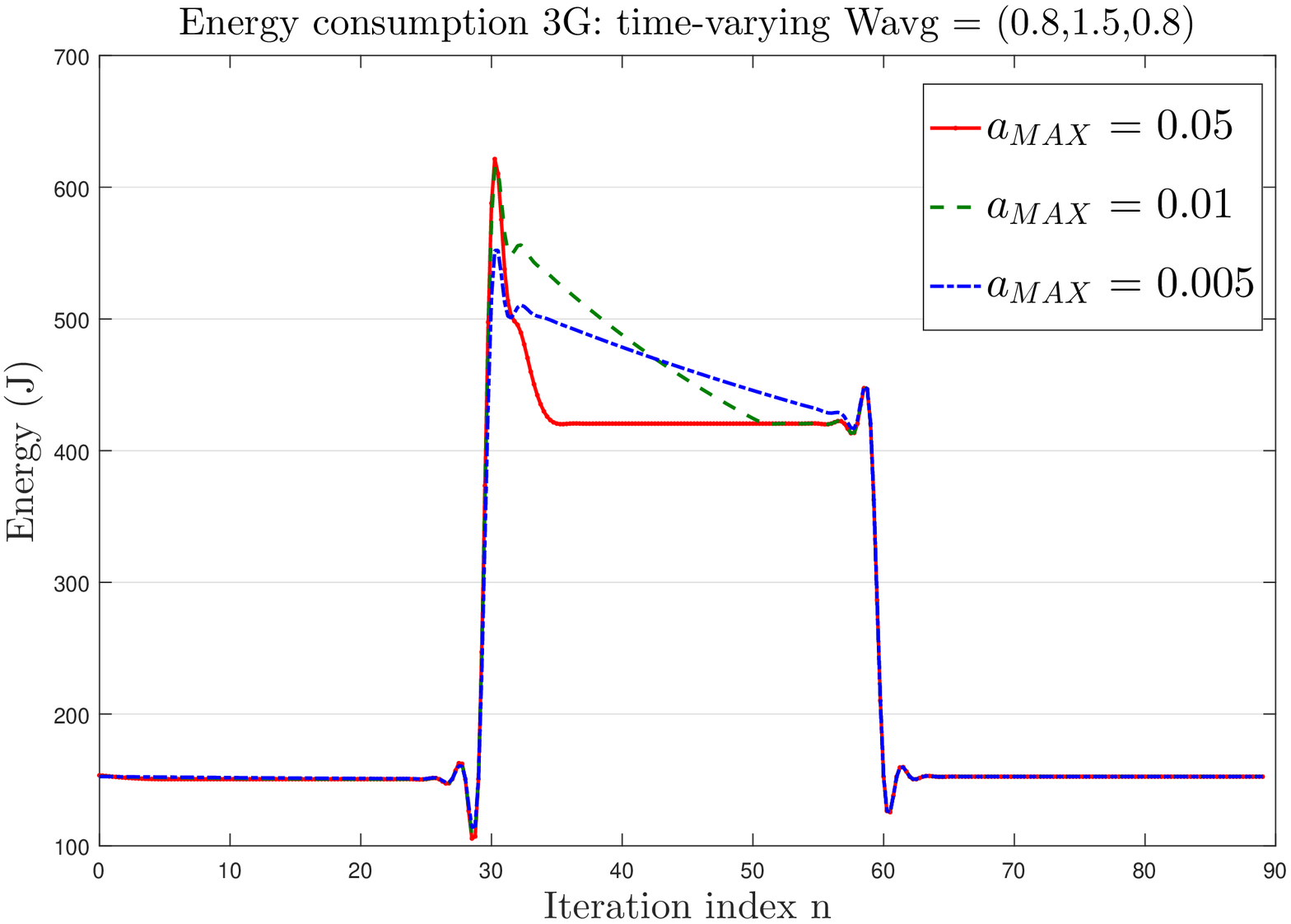}
}
\subfloat[{$\overline{w} = [11.25, 24, 11.25]$} ]{\includegraphics[width=.6\columnwidth]{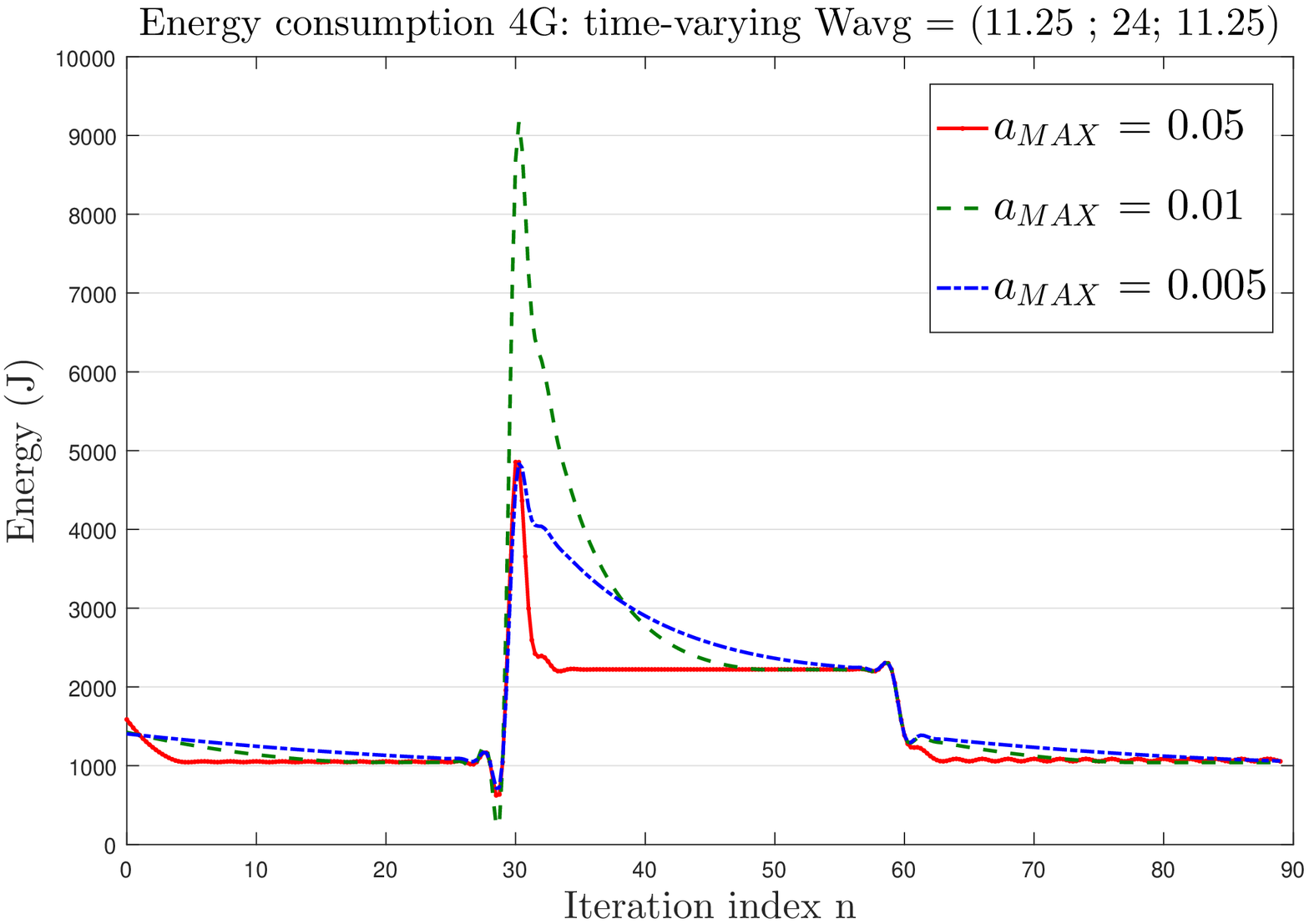}
}%
\subfloat[{$\overline{w} = [4, 8, 4]$} ]{\includegraphics[width=.6\columnwidth]{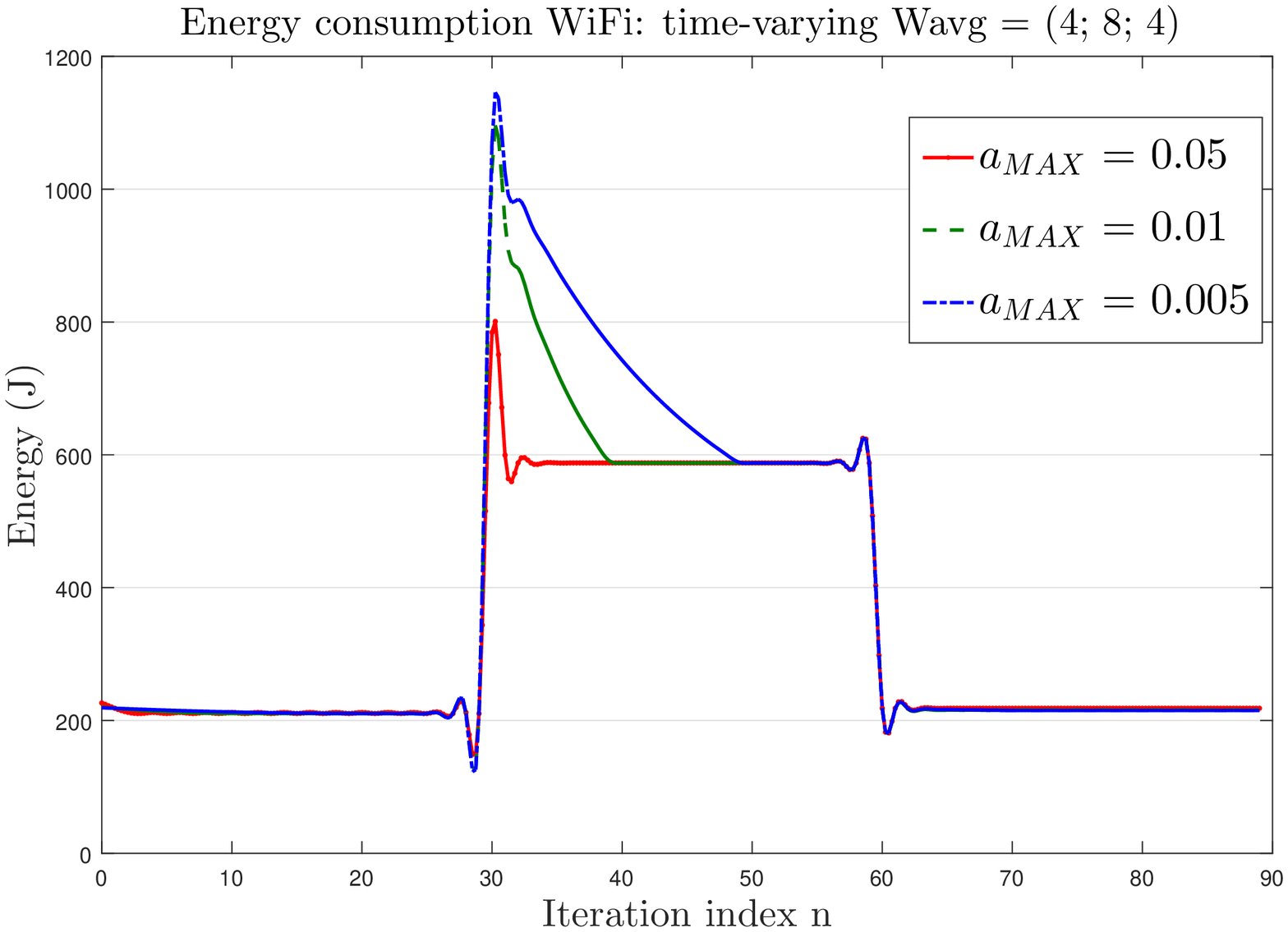}
} 
\vspace{-8pt}
\caption[Time evolutions (in the n index) of the energy consumption of the proposed bandwidth manager, case of time-varying $\overline{w}$.]{ Time evolutions (in the n index) of the energy consumption of the proposed bandwidth manager, case of time-varying $\overline{w}$, at: (a)\: $\widehat{R} = 1.8 \:(Mb/s),\: M_{0} = 256\: (Mb), \: \beta = 2, \: \Delta_{TM} = 1460\:(s), \: \Delta_{DT} = 0.14\: (s),$ for 3G scenario; (b) \: $\widehat{R} = 45 \:(Mb/s),\: M_{0} = 256\: (Mb), \: \beta = 2.33, \: \Delta_{TM} = 58.6\:(s), \: \Delta_{DT} = 5.61\times10^{-3}\: (s),$ for 4G scenario; (c) \: $\widehat{R} = 9.9 \:(Mb/s),\: M_{0} = 256\: (Mb), \: \beta = 2.33, \: \Delta_{TM} = 266\:(s), \: \Delta_{DT} = 2.55\times10^{-2}\: (s),$ for WiFi scenario. }
\label{fig:4.5}
\end{figure*}

As pointed out in \cite{19a_xu2014managing}, memory contention phenomena and/or network congestions may produce abrupt (typically, unpredictable) time-variations of the parameters $\overline{w}$ and or $K_{0}$ 
Hence, in order to evaluate the tracking capabilities of the proposed adaptive bandwidth manager 
and its sensitivity to the parameters $a_{MAX}$ 
in Fig. \ref{fig:4.5}, we report the measured behaviors of the energy sequence: $\{ \mathcal{E}_{TOT}^{\ast(n)}, \:\: n \geq 0\}$ when, due to memory contention phenomena, the memory dirty rate $\overline{w}$ of the running \emph{memtester} application abruptly varies. 

\begin{figure*}[]
\centering
\subfloat[{$K_0 = [0.18, 18, 0.18]$}]{\includegraphics[width=.6\columnwidth]{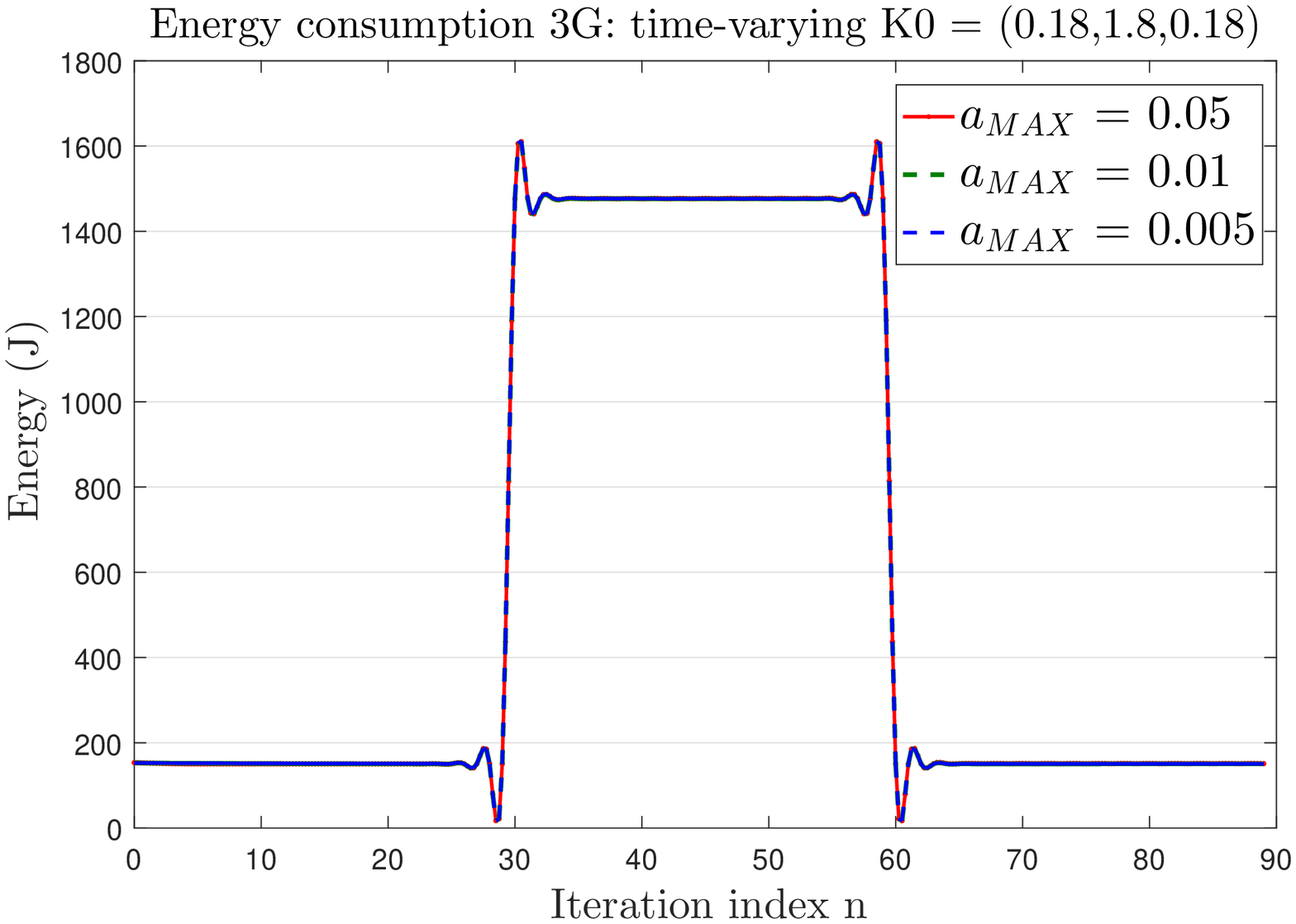}%
}
\subfloat[{$K_0 = [0.09, 0.9, 0.09]$}]{\includegraphics[width=.6\columnwidth]{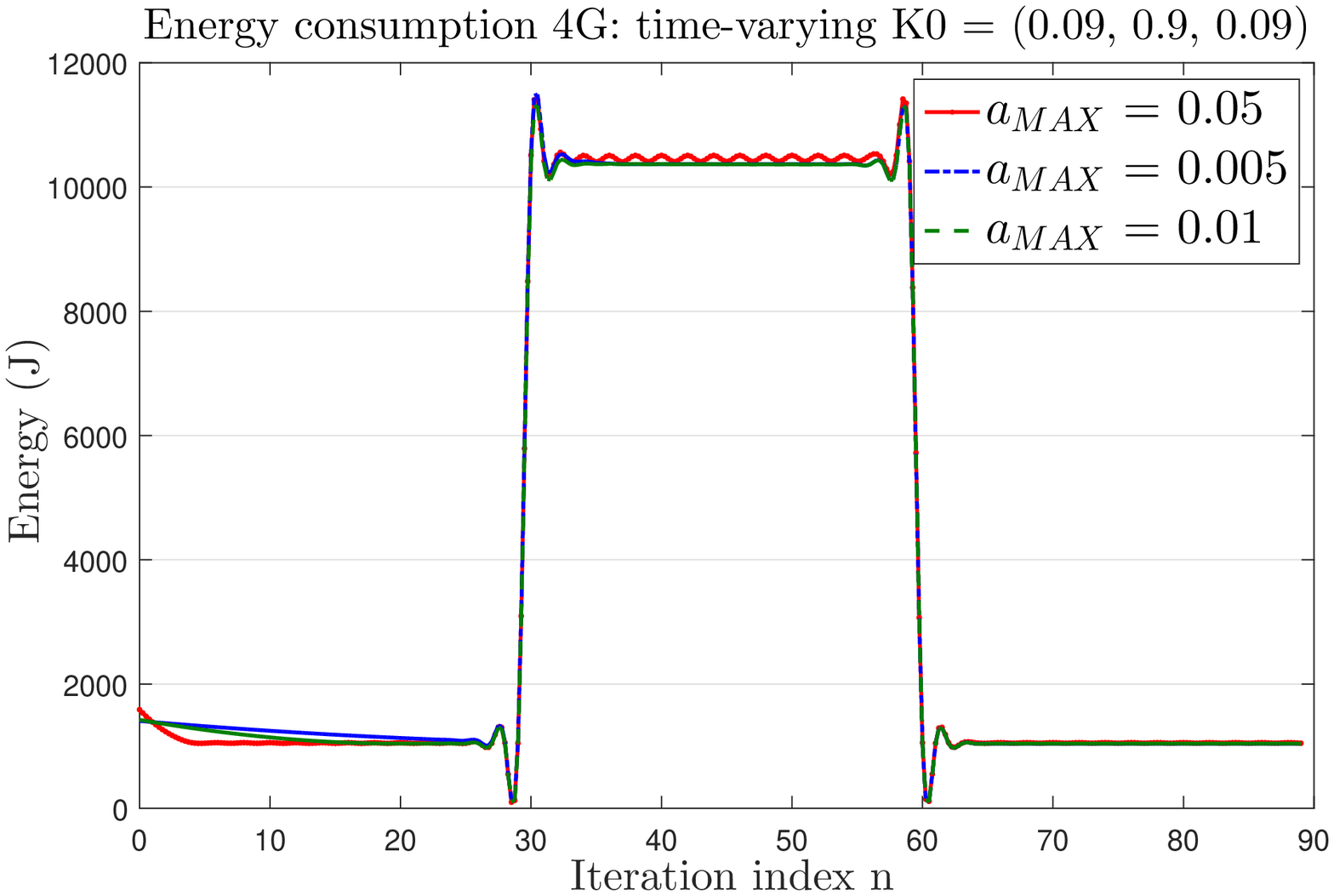}%
}
\subfloat[{$K_0 = [0.05, 0.5, 0.05]$}]{%
\includegraphics[width=.6\columnwidth]{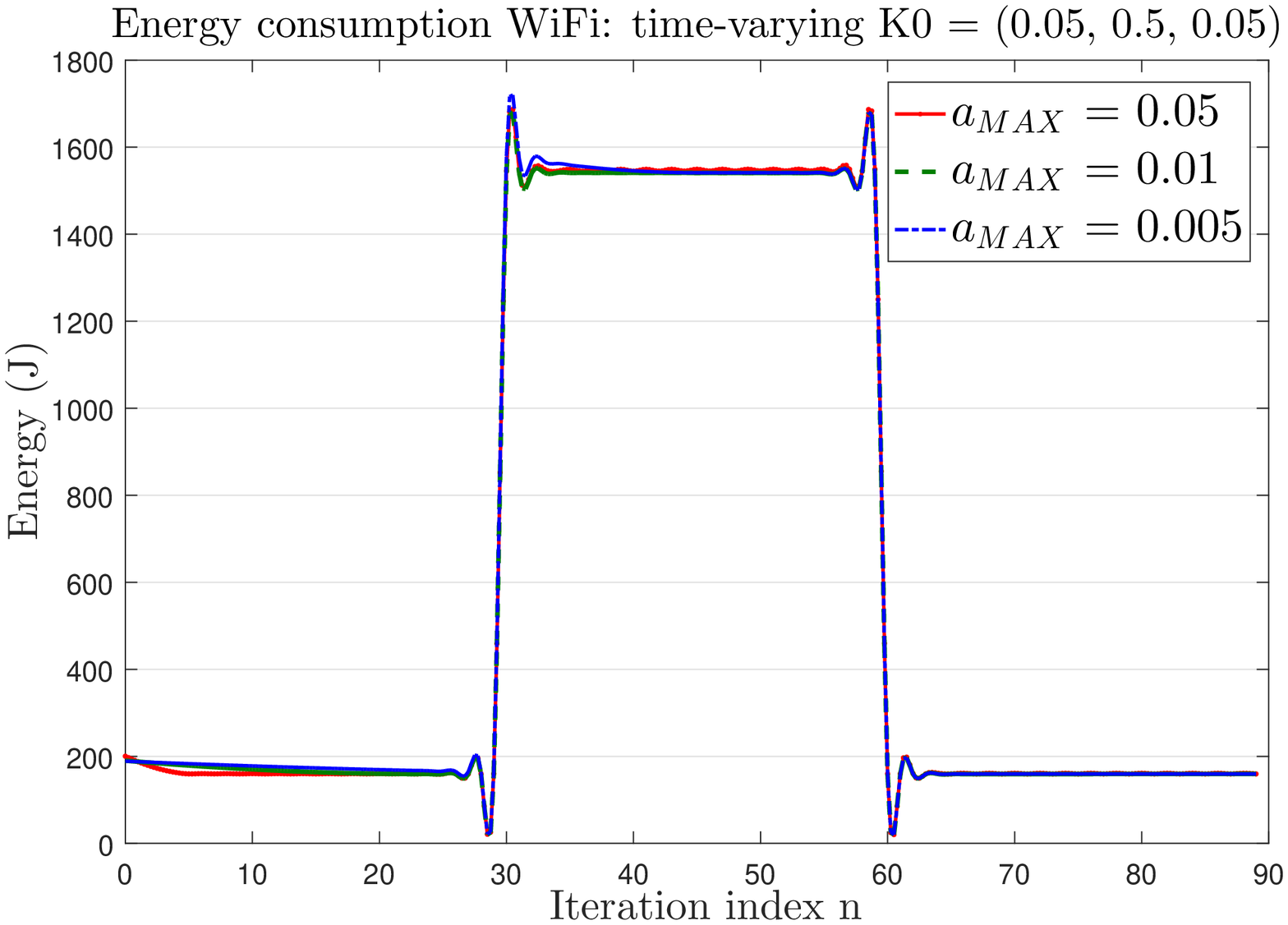}
} 
\caption[Time evolutions (in the n index) of the energy consumption of the proposed bandwidth manager, case of time-varying $K_{0}$.]{ Time evolutions (in the n index) of the energy consumption of the proposed bandwidth manager, case of time-varying $K_{0}$, at: (a)\: $\widehat{R} = 1.8 \:(Mb/s),\: M_{0} = 256\: (Mb), \: \beta = 2, \: \Delta_{TM} = 1460\:(s), \: \Delta_{DT} = 0.14\: (s),$ for 3G scenario; (b) \: $\widehat{R} = 45 \:(Mb/s),\: M_{0} = 256\: (Mb), \: \beta = 2.33, \: \Delta_{TM} = 58.6\:(s), \: \Delta_{DT} = 5.61\times10^{-3}\: (s),$ for 4G scenario; (c) \: $\widehat{R} = 9.9 \:(Mb/s),\: M_{0} = 256\: (Mb), \: \beta = 2.33, \: \Delta_{TM} = 266\:(s), \: \Delta_{DT} = 2.55\times10^{-2}\: (s),$ for WiFi scenario. }
\label{fig:4.6}
\end{figure*}

An examination to the plots of Fig. \ref{fig:4.5} and Fig. \ref{fig:4.6} supports three main conclusions. 
\begin{itemize}
	\item First, according to the fact that the energy function increases for increasing $\overline{w}$ and/or $K_{0}$, all the plots of Fig. \ref{fig:4.5} and Fig. \ref{fig:4.6} scale up at n = 30 and, then, scale down at n = 60.
	\item Second, the proposed bandwidth manager quickly reacts to abrupt unpredicted time variations of the migrating application and/or underlying network connections.
	\item Third, while virtually indistinguishable plots are obtained for $a_{MAX}$ ranging over the interval $[5\times10^{-2} , 5\times10^{-3}]$ in case of time-varying $K_{0}$ (see. Fig. \ref{fig:4.6}), the same results is not obtained in case of time-varying $\overline{w}$ (see. Fig. \ref{fig:4.6}). This phenomenon is due to the fact that while $K_{0}$ is a multiplicative constant in the formula of the energy, $\overline{w}$, in addition to a profound impact on energy, causes that our TCBM uses more iterations to go from transient-states to the steady-states. Precisely, it is showed that, the decrease of $a_{MAX}$ increases the number of iterations that are used by the software to return to the equilibrium state.
\end{itemize}

For this reason we prefer to work with $a_{MAX}$ high, over the interval [0.5 , 0.05], in such a way that (in a maximum of six or seven iterations), the software reacts well to variations of $\overline{w}$. 

Overall, from the outset, we conclude that the proposed adaptive bandwidth manager is robust with respect to the actual tuning of $a_{MAX}$, at least for values of $a_{MAX}$ ranging over the the interval [0.5 , 0.05], in  order to exhibits the best trade-off among the contrasting requirements of short transient-states and stable steady-states.

\subsection{Comparative energy tests under random migration ordering and synthetic workload}\label{sez:4.5}
The benchmark bandwidth management policy of the Xen hypervisor 
does not guarantee, by design, minimum energy consumptions and does not enforce QoS constraints on the resulting memory migration and stop-and-copy times. Furthermore, differently from  $\tilde{I}_{MAX}$, 
the maximum number of allowed rounds: $I_{MAX}^{XEN}$ is fixed by the Xen hypervisor in an application-oblivious way (typically, $I_{MAX}^{XEN} \leq 29$; see \cite{42a_chisnall2008definitive,46a_hwang2013distributed}). 
Hence, in order to carry out \emph{fair} energy comparisons, in the carried out field trials, we proceed as follows:
\begin{itemize}
\item[i)] set $I_{MAX}^{XEN}$ and $R_{MAX}^{XEN}$;
\item[ii)] measure the resulting Xen energy consumption $\mathcal{E}_{TOT}^{XEN}$, speed-up factor $\beta^{XEN}$,  total migration time $T_{TM}^{XEN}$, downtime $T_{DT}^{XEN}$;
\item[iii)] enforce $\widehat{R} \equiv R_{MAX}^{XEN}$, together with the QoS constraints: $\Delta_{TM} \equiv T_{TM}^{XEN}$,  $\Delta_{DT} \equiv T_{DT}^{XEN}$, and $\beta \equiv \beta^{XEN}$;
\item[iv)] measure the resulting energy consumption $\mathcal{E}_{TOT}^{\ast}$ of the proposed bandwidth manager at $I_{MAX} = \tilde{I}_{MAX}$. %
\end{itemize}

The \textit{memtester} (see in \cite{Baccarelli20151})
is the application considered in this section and the implemented migration ordering of the dirtied memory pages is the random one.

The numerical results measured through a campaign of trials developed for the three considered scenarios (3G, 4G and WiFi) are partially omitted cause the lack of space. We show only the table data referred top the 4G scenario. %

\begin{table}[h]
\caption{Scenario 4G with $ M_{0} = 256 (Mb);\:\: \alpha = 2; \:\: K_{0} = 0.09; \:\: E_{SETUP} = 5.1 (J);$  (a) $\biggl (\dfrac{\overline{w}}{ \widehat{R}}\biggl) = 0.33$ and $\widehat{R} = 0.33 \times R_{MAX}^{XEN} = 14.85 (Mb/s);$  (b) $\biggl (\dfrac{\overline{w}}{ \widehat{R}}\biggl) = 0.11$ and $\widehat{R} =  0.11 \times R_{MAX}^{XEN} = 4.95 (Mb/s);$}
\centering
\label{table:2} 
\begin{tabular}{llll}
\toprule
$I_{MAX}^{XEN}$ & 6  & 14 & 25\\
\midrule
$T_{DT}^{XEN} = \Delta_{DT}$(s)   & 0.103 &  $5.42 \times 10^{-4}$ &  $4.03 \times 10^{-7}$\\ \\
$T_{TM}^{XEN} = \Delta_{TM}$(s)   & 46.9 &  65.2 &  83.6\\ \\
$\beta$   & 1.87 &  1.95 &  1.98\\ \\
$\mathcal{E}_{TOT}^{XEN}$(J)   & 1880 &  2150 &  2470\\ \\
$\mathcal{E}_{TOT}^{LIV\_MIG}$(J)   & 1550 &  1550 &  1550\\\\
Q   & 1 &  1 &  1\\ \\
$\mathcal{E}_{TOT}^{TCBM}$(J)   & 1366 &  1373 &  1373\\ \\
En. save vs. XEN (\%) & 27.3 & 36.1 &44.4 \\ \\
En. save vs. LIV\_MIG (\%) & 11.8 & 11.4 &11.4 \\ \\
(a)\\
\bottomrule \\   
\end{tabular}
\hspace*{-3pt}\makebox[\linewidth][c]{%
\begin{tabular}{llll}
\toprule
$I_{MAX}^{XEN}$ & 6  & 14 & 25\\
\midrule
$T_{DT}^{XEN} = \Delta_{DT}$(s)   & $5.9 \times 10^{-4}$ &  $4.07 \times 10^{-9}$ &  $3.4 \times 10^{-16}$\\ \\
$T_{TM}^{XEN} = \Delta_{TM}$(s)   & 84.9 &  110 &  137\\ \\
$\beta$   & 4.47 &  4.78 &  4.89\\ \\
$\mathcal{E}_{TOT}^{XEN}$(J)   & 632 &  624 &  602\\ \\
$\mathcal{E}_{TOT}^{LIV\_MIG}$(J)  & 1170 &  1170 &  1170\\\\
Q   & 1 &  1 &  1\\ \\
$\mathcal{E}_{TOT}^{TCBM}$(J)   & 531.7 &  545.25 &  541.8\\ \\
En. save vs. XEN \small{(\%)} & 15.8 & 12.6 &10 \\ \\
En. save vs. LIV\_MIG \small{(\%)} & 54.5 & 53.4 &53.7 \\\\
(b)\\
\bottomrule     
\end{tabular}}
\end{table}
	
These table show the energy values obtained using Xen, the bandwidth management policy developed in the paper \cite{Baccarelli20151}, and the Tunable-complexity bandwidth manager.%

An examination of the results of data leads to two main conclusion. 
First, in all the carried out field trials the percent energy saving: 
\begin{itemize}
\item $(1 - (\mathcal{E}_{TOT}^{*} / \mathcal{E}_{TOT}^{XEN}))$\% of the proposed bandwidth manager over the Xen one is between  3\% (minimum value of energy saving) for $(\overline{w}/\widehat{R}) = 0.11$ and $I_{MAX} = 25$, to 44.4\% (maximum value of energy saving) for $(\overline{w}/\widehat{R}) = 0.33$ and $I_{MAX} = 25$ (see Table \ref{table:2}(a));
\item $(1 - (\mathcal{E}_{TOT}^{*} / \mathcal{E}_{TOT}^{LIV_MIG}))$\% of the proposed bandwidth manager over the BMOP (Bandwidth Management Optimization Problem, see paper \cite{Baccarelli20151}) is between  11.2\% (minimum value of energy saving) for $(\overline{w}/\widehat{R}) = 0.33$ and $I_{MAX} = 6$, to 54.5\% (maximum value of energy saving) for $(\overline{w}/\widehat{R}) = 0.11$ and $I_{MAX} = 6$ (see Table \ref{table:2}(b)).
\end{itemize}

In all scenarios, TCBM appears to be the best one from the point of view of energy saving. These noticeable energy gains support the conclusion that the bandwidth management policy developed in this paper is the optimal one, and, by design, it minimizes the migration-induced energy consumption.

Second, the values of the measured energy gains mainly depend on the considered ratio: $(\overline{w} / \widehat{R})$. In particular, in these tests only values of $(\overline{w} / \widehat{R}) \leq 0.33$ are considered, because, if and only if this constraint is satisfies, the Xen (heuristic) bandwidth management policy presents decreasing values of energy for increasing values of $I_{MAX}$. Hence, under this condition, it make sense to compare our bandwidth manager with Xen and BMOP. 

In the carried out tests, is reported that, while the TCBM in each scenario presents a constant gain with respect to the optimization method described in \cite{Baccarelli20151}, from the comparison with Xen comes out that the percentage of energy saving tends to decrease (for increase of $I_{MAX}$)
when the ratio $(\overline{w} / \widehat{R}) < 0.33$; on the contrary the percentage of energy saving tends to increase when the ratio $(\overline{w} / \widehat{R}) = 0.33$. 

In all the experiments, $R_{MAX}$ was chosen equal to the value of $R_{MAX}$ of 3G (which turns out to be smaller, than those in the 4G and WiFi), in such a way to have the comparisons in a consistent manner.

\begin{figure*}[]
\centering
\subfloat[3G Scenario]
{%
\includegraphics[width=.6\columnwidth]{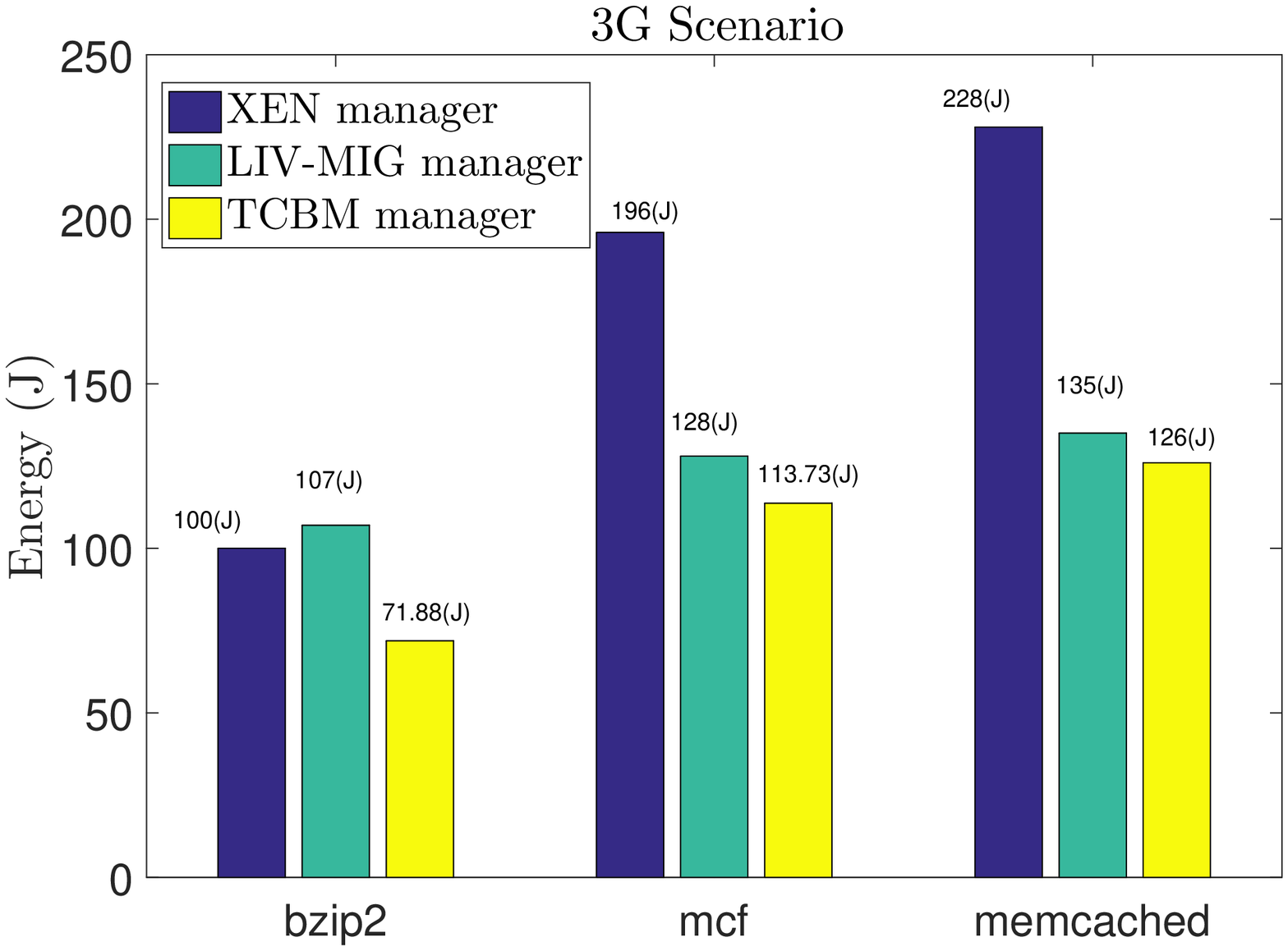}%
} 
\subfloat[4G Scenario]
{%
\includegraphics[width=.6\columnwidth]{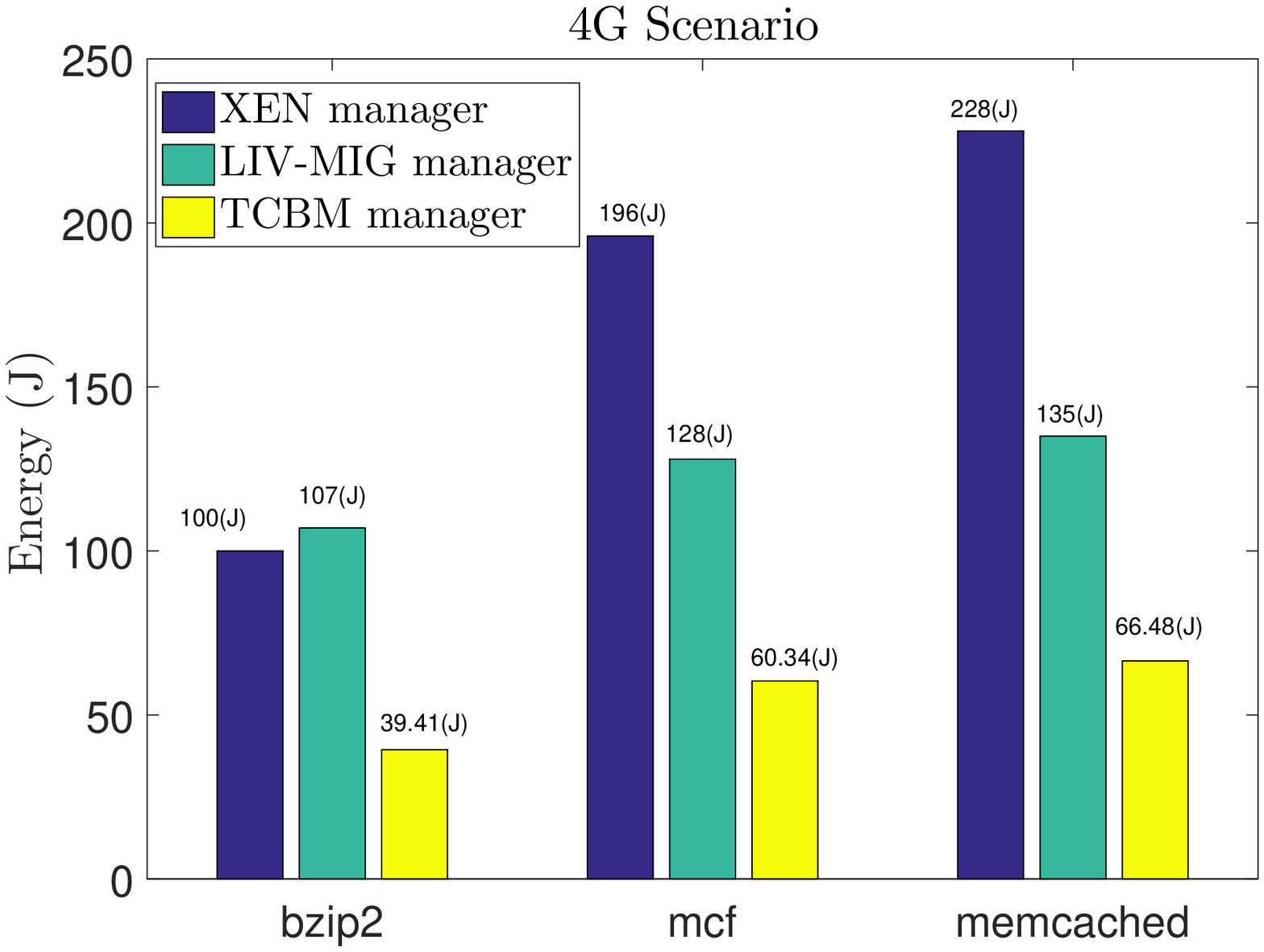}%
} 
\subfloat[WiFi Scenario]
{%
\includegraphics[width=.6\columnwidth]{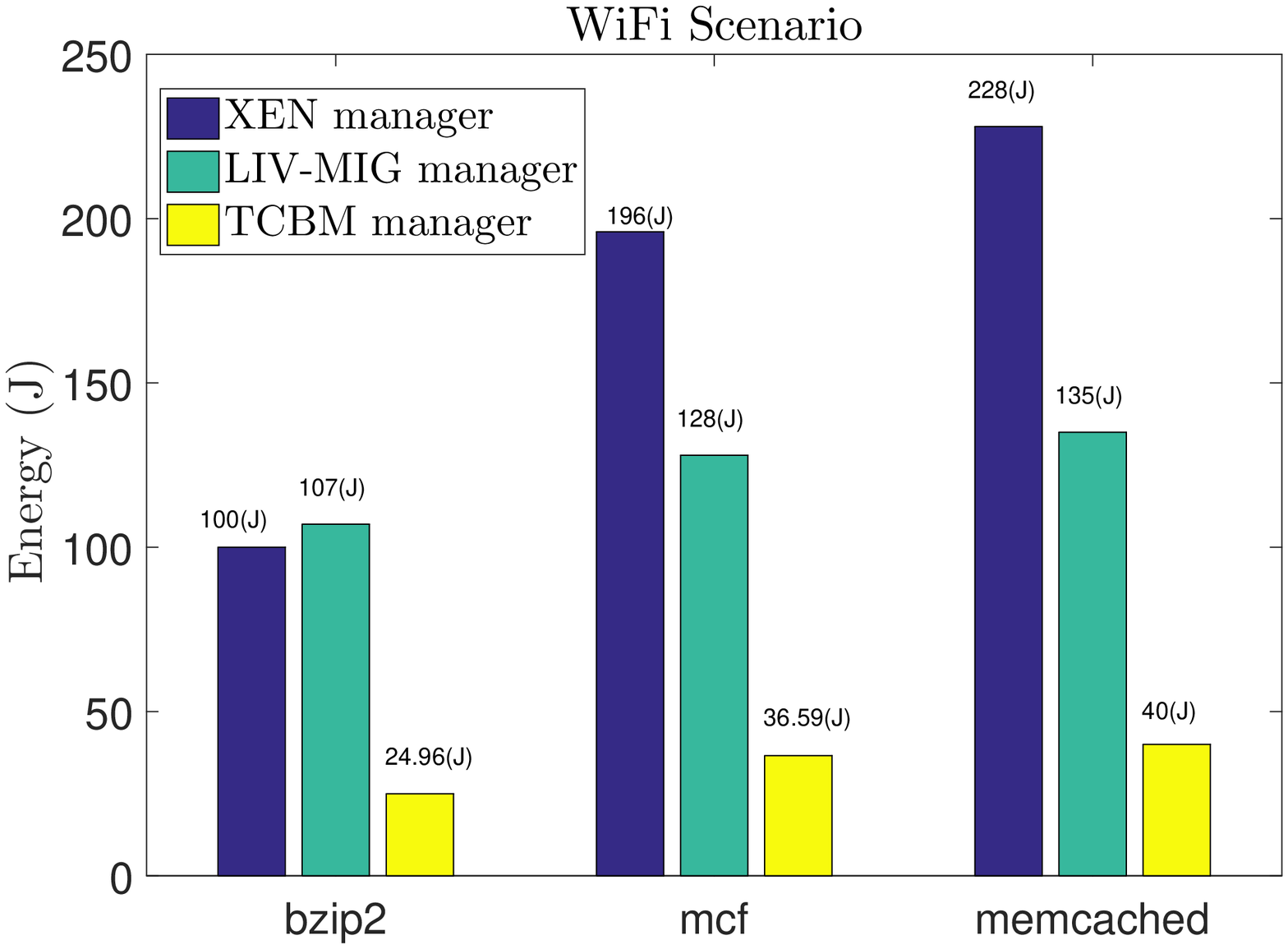}%
} 
\vspace{-8pt}
\caption{Energy consumptions obtained by \textit{bzip2}, \textit{mcf} and \textit{memcached} in : (a) 3G scenario; (b) 4G scenario; (c) WiFi scenario.}
\label{fig:4.7}
\end{figure*}

The Figures \ref{fig:4.7} show the results of the tests.	 

An examination of the bar plots of Fig.\ref{fig:4.7} leads to two main conclusion. First, since the dirty rate increases by passing from the (read-intensive) \emph{bzip2} program to the (write-intensive) \emph{memcached} one, the corresponding energy consumptions also exhibit increasing trends under both the Xen, LIV-MIG \cite{Baccarelli20151} and proposed bandwidth managers. Second, in all cases, the energy consumption relating to the migration by applying our method appears to be lower than both Xen and LIV-MIG manager. In particular, the percent energy savings of the proposed manager over the Xen and the LIV-MIG under the \emph{bzip2}, \emph{mcf} and \emph{memcached}, for each application scenarios are reported in Table \ref{table:2} and Table \ref{table:3}.

\begin{table}[h!]
\small
\caption{Percent energy savings of the TCBM manager over Xen and LIV-MIG managers.} 
\label{table:3} 
\centering
\hspace*{-3pt}\makebox[\linewidth][c]{%
\begin{tabular}{|c|c|c|c|c|}
\hline
\textbf{}& \textbf{Parameter} & \textbf{\emph{bzip2}} & \textbf{\emph{mcf}} & \textbf{\emph{memc.}}\\
\midrule
\multirow{2}{*}{3G}   &Energy saving resp. Xen(\%) & 28.1 & 41.92 &44.74\\
\cline{2-5}
& 	En. saving resp. LIV-MIG(\%)  &32.8 &11.15&6.67 \\
\cline{1-5}
\midrule   
\multirow{2}{*}{4G}   &Energy saving resp. Xen(\%) & 60.5 & 69.21 &70.84\\
\cline{2-5}
& 	En. saving respect LIV-MIG(\%)  &63.17&52.86&50.76 \\
\cline{1-5}
\midrule 
\multirow{2}{*}{WiFi}   &Energy saving resp. Xen(\%) & 75.04& 81.33 &82.46\\
\cline{2-5}
& 	En. saving resp. LIV-MIG(\%)  &76.67 &71.41&70.37 \\
\cline{1-5}
\bottomrule  
\end{tabular}
}
\end{table}

This confirms the trend of the previous Section \ref{sez:4.5} about the large energy-gains offered by the proposed manager under write-intensive applications.

\section{Conclusion}
\label{sec:conclusion}
In this paper we presented a novel approach for bandwidth management in live migration virtual machine in wireless context. 
Our results show a significant improvement with respect to the currently used approach in most relevant implementation architecture for live virtual machines.


%
%



\bibliographystyle{IEEEtran}
\bibliography{bibliography} 
%
%
%

\end{document}